\documentclass[12pt]{article}
\usepackage{jheppub}
\usepackage{comment}

\title{Excited States of Open Strings From $\Ncal=4$ SYM}
\author[a]{Eric Dzienkowski}
\affiliation[a]{Department of Physics, University of California, Santa Barbara, CA 93106}
\emailAdd{eric.m.dzienkowski@gmail.com}
\date{\today}

\abstract{%
We continue the analysis of building open strings stretched between giant gravitons from $\Ncal=4$ SYM by going to second order in perturbation theory using the three-loop dilatation generator from the field theory.
In the process we build a Fock-like space of states using Cuntz oscillators which can be used to access the excited open string states.
We find a remarkable cancellation among the excited states that shows the ground state energy is consistent with a fully relativistic dispersion relation.}

\begin{document}

\allowdisplaybreaks

\newcommand{\Ncal}{\mathcal{N}}
\newcommand{\Ocal}{\mathcal{O}}
\newcommand{\Pcal}{\mathcal{P}}
\newcommand{\zbb}{\bar{z}}
\newcommand{\xibb}{\bar{\xi}}
\newcommand{\xitt}{\tilde{\xi}}
\newcommand{\Jhh}{\hat{J}}
\newcommand{\etabb}{\bar{\eta}}
\newcommand{\etatt}{\tilde{\eta}}
\newcommand{\Ych}{\check{Y}}
\newcommand{\Zch}{\check{Z}}
\newcommand{\sutwo}{\mathfrak{su}(2)}
\newcommand{\kgs}{\ket{\Omega^{(0)}}}
\newcommand{\bgs}{\bra{\Omega^{(0)}}}
\newcommand{\Arg}{\text{Arg}}
\newcommand{\bbr}{\mathbb{R}}

\newcommand{\bra}[1]{\left\langle #1 \right|}
\newcommand{\ket}[1]{\left| #1 \right\rangle}
\newcommand{\braket}[2]{\left\langle #1 \middle| #2 \right\rangle}
\newcommand{\exval}[3]{\left\langle #1 \middle| #2 \middle| #3 \right\rangle}
\newcommand{\pderiv}[2]{\frac{\partial #1}{\partial #2}}

\newcommand{\beq}{\begin{equation}}
\newcommand{\eeq}{\end{equation}}
\newcommand{\Tr}{\text{Tr}}

\newcommand{\lp}{\left(}
\newcommand{\rp}{\right)}
\newcommand{\lr}{\left.}
\newcommand{\lb}{\left[}
\newcommand{\rb}{\right]}
\newcommand{\rr}{\right.}
\newcommand{\ls}{\left|}
\newcommand{\rs}{\right|}

\newcommand{\rgg}{R_{\text{GG}}}

\renewcommand{\det}{\text{det}}

\newcommand{\idea}[1]{{\color{blue}\em #1}}
\newcommand{\checkthis}[1]{{\color{red}\bf #1}}
\newcommand{\needcite}{{\color{green} need citation}}

\maketitle


\section{Introduction}

String theory to date has no non-perturbative defintion.
The best we can do is through gauge / gravity duality and more generally, holography.
The AdS / CFT correspondence \cite{Maldacena:1997re} provides a concrete example of these concepts.
In particular we can define a specific class of quantum type IIB string theories, those embedded in spacetimes which are asymptotically $AdS_5\times S^5$ with background five form flux, in terms of the very special superconformal four-dimensional field theory $\Ncal=4$ SYM on $\mathbb{R}\times S^3$ with gauge group $SU(N)$ ($U(N)$).
The definition is encoded by a one-to-one and onto mapping between quantum strings and quantum operators of the gauge theory.
In the mapping we should find all the objects known to string theory, perturbative and non-perturbative.
In addition to closed strings, we should be able to find open strings and even D-branes living in the gauge theory.

The quantization of the string in $AdS_5\times S^5$ is a non-trivial problem due to the complexity and non-linearity of the equations of motion of the sigma model \cite{Metsaev:1998it}.
However, the integrability of the string sigma model \cite{Bena:2003wd} becomes a powerful tool by which to solve the model and access the spectrum of closed string states.
The string sigma model has been solved for a large class of closed strings in terms of algebraic curves in a complex manifold \cite{Beisert:2005bm}.
Computation of the quantum fluctuations of algebraic curves provides information about the excited closed string states.

The spectrum of $\Ncal=4$ SYM is given by the eigenvalues of the dilatation operator if we put the theory on $\bbr^4$ instead of $\bbr\times S^3$.
In general this is incredibly difficult, but simplifies in the planar limit.
Planar $\Ncal=4$ SYM is integrable up to one-loop \cite{Beisert:2003yb} and expected to be integrable to all orders in perturbation theory.
The spectrum of anomalous dimensions is characterized by a Bethe ansatz for an integrable $\mathfrak{psu}(2,2|4)$ spin chain \cite{Beisert:2005fw}.
A major success of the integrability program which relates the string theory to the gauge theory is seen in the thermodynamic limit, where the solutions to the Bethe ansatz become the algebraic curves of the closed string \cite{Beisert:2005di}.
Consequently, we should identify the planar limit of $\Ncal = 4$ SYM with classical string theory on $AdS_5\times S^5$.

It is natural to ask whether any of this integrability carries over to the open string sector of the correspondence.
The string theory is integrable for some sets of boundary conditions \cite{Mann:2006rh}, but for generic boundary conditions integrability is not expected hold.
In $\Ncal = 4$ SYM one can use heavy operators, e.g. determinants and subdeterminants, to impose boundary conditions on the spin chains.
These heavy operators are dual to giant gravitons in the string theory \cite{McGreevy:2000cw}.
For maximal giants the open spin chain system may be integrable \cite{Berenstein:2006qk} but for non-maximal giants the boundary Yang-Baxter equations are not satisfied \cite{Koch:2015pga} which indicates that integrability may not be preserved.
The integrability of the open string remains an open question, and if it is not there one wonders if there are other structures that replace it.

Without a tool as powerful as integrability in the open string sector, we must compute the spectrum of open strings by more traditional means.
On the string side, we can solve the equations of motion and try to quantize quadratic fluctuations.
On the gauge theory side, we need to find the dual operators and compute their  anomalous dimensions.
Although the traditional approach is more cumbersome, it has its advantages.
Because integrability is so powerful, it sometimes hides the underlying dynamics that allow for cancellations to occur and for results to be so simply represented.
Working directly with field theories means we are purposefully not hiding from the dynamics.
The physical description of these results remains very apparent which is sometimes obfuscated by the mathematical structure of integrability.
At the same time, we would still like to know if there is a mathematical structure analogous to integrability in the open string sector that unifies and simplifies some of those calculations.
One way of answering this question is to take a step back and continue to work in a more traditional realm.

We work with open string states constructed in the $\sutwo$ sector of the gauge theory, continuing the work of \cite{Berenstein:2013eya}.
The $\sutwo$ sector is made up of single and multi-trace operators constructed from two of the complex scalars, $Y$ and $Z$.
This is the smallest non-trivial compact subsector of $\Ncal=4$ SYM closed under renormalization \cite{Beisert:2003tq}.
The class of D-branes known as giant gravitons was constructed in this subsector \cite{Balasubramanian:2001nh}.
Means by which to build open strings stretched between these D-branes was suggested in \cite{Balasubramanian:2002sa}.
The properties of this class of D-branes are very important as they determine a set of boundary conditions for the sigma model.
When open string excitations are built explicitly from the $Y$ and $Z$ fields, the operators lie in the $\sutwo$ sector of the gauge theory.
One can recast the problem of anomalous dimensions for these open string states in terms of Cuntz oscillators \cite{Berenstein:2005fa}.
The Hamiltonian for this system was computed at two-loops in \cite{Berenstein:2013eya}.
In this work, the open strings were repackaged as Cuntz oscillator coherent states.
The ground state at one-loop and the first order correction to the energy using the two-loop Hamiltonian were computed.
Together with the tree level result, the energy of the ground state was consistent with a relativistic dispersion relation up to second order in the 't Hooft coupling.
It was argued that these strings were probing the geometry of the $S^3$ worldvolume of the giant gravitons directly.

Operators in the $\sutwo$ sector are dual to strings propagating in an $\bbr_t\times S^3$ background of the $AdS_5\times S^5$ where $\bbr_t$ is the time direction in $AdS_5$ and $S^3\subset S^5$.
A large class of closed string solutions to the equations of motion in this subsector have been found including circular strings, folded strings, spiky strings, and solitonic strings \cite{Frolov:2003qc,Plefka:2005bk,Kruczenski:2006pk,Chen:2006gea}.
The solitonic strings are called giant magnons \cite{Hofman:2006xt}.
Bound states of many giant magnons have been found in the gauge theory by studying the asymptotic S-matrix \cite{Dorey:2006dq}.
By truncating the giant magnon solution, one can obtain an open string solution with the boundary condition that the string ends on two giant gravitons \cite{Berenstein:2014isa}.
This string solution satsifies the same dispersion relation conjectured for the open string coherent states in the gauge theory.
Furthermore, these two solutions have the same internal structure; they both can be characterized by linear interpolations between the giant gravitons in the complex unit disk.
It is thus natural to identify these object under the AdS / CFT duality.

In this note we explore the excited states of this class of open strings from the gauge theory side.
Even without integrability, we find regions of the problem space that are tractable.
In particular, we are able to compute the next order corrections to the ground state in perturbation theory exactly.
Perturbation theory at second order is non-trivial in this case for two reasons.
The first is that one must sum over the entire spectrum of the lowest order Hamiltonian.
The fact that the energy can be computed analytically reveals a marvelous cancellation among the excited energies, despite not being able to compute them exactly.
The second is that we will need the three-loop dilatation operator in the $\sutwo$ sector.
The higher-loop contribution provides a necessary cancellation to realize the open string duality.
This shows that the dilatation operators at each loop order are related to each other in a very special way, even in the presence of open boundary conditions.
The second order correction to the energy is shown to be consistent with the relativistic dispersion relation conjectured in \cite{Berenstein:2013eya}.

This note is organized as follows.
In Section \ref{sec:openstrings} we review the construction of giant gravitons and open strings stretched between them in the gauge theory.
In Section \ref{sec:dilatation} we review how to construct the relevant open strings from the gauge theory using the dilatation generator to govern its dynamics. 
In Section \ref{sec:excited} we construct a Fock-like space of states and discuss the excited states with respect to the one-loop dilatation generator.
In Section \ref{sec:ptheory} we solve the first order correction to the ground state energy and the second order correction to the ground state energy showing consistency with a relativistic dispersion relation.
We then conclude with future directions in Section \ref{sec:conclusions}.


\section{Constructing Open Strings}
\label{sec:openstrings}

D-branes are non-perturbative objects in string theory lacking a full quantum description.
Calculations involving D-branes come in the form as BPS solutions to the supergravity equations of motion, classical analysis of the DBI action, or perturbatively in the form of open strings.
Using $\Ncal = 4$ SYM as a definition of quantum string theory provides the best approach to formulating D-branes non-perturbatively.
The class of D-branes known giant gravitons \cite{McGreevy:2000cw} has been successfully constructed in the gauge theory.
The particular giant gravitons we are interested in are D-branes with the world-volume of an $S^3$ traversing a great circle of the $S^5$ in $AdS_5\times S^5$.
The radius of the giant graviton is proportional to its velocity, or to its angular momentum around the $S^5$.
They are half BPS whose angular momentum has an upper bound due to the fact that the radius of an $S^3$ subspace of $S^5$ is bounded above by that of the $S^5$.
The dual operators in the gauge theory were found in \cite{Balasubramanian:2001nh} and determined to be subdeterminant operators of the form
\beq
\label{eq:subdet}
\det_k(Z) = \frac{1}{N!}{N\choose k} \epsilon^{a_1\dots a_k c_{k+1}\dots c_N}\epsilon_{b_1\dots b_k c_{k+1}\dots c_N} Z_{a_1}^{b_1}\cdots Z_{a_k}^{b_k}
\eeq
When $k=N$ we just have $\det_N(Z) = \det(Z)$ which is called a maximal giant.
The classical upper bound on the angular momenta is directly related to the condition that $k\leq N$.

One can build string excitations, represented as operators $W_i$, on top of maximal giants by contracting the indices of the epsilon symbols with additional operators instead of with the identity.
\beq
\Ocal(W_1,\dots,W_s) = \frac{1}{N!}{N \choose s}\epsilon^{a_1\dots a_{N-s} c_{1}\dots c_s}\epsilon_{b_1\dots b_{N-s} d_{1}\dots d_s} Z_{a_1}^{b_1}\cdots Z_{a_{N-s}}^{b_{N-s}} {W_1}_{c_1}^{d_1}\cdots {W_s}_{c_s}^{d_s}
\eeq
These operators are given by the generating function equation
\beq
\label{eq:genfunc}
\Ocal(W_1,\dots,W_s) = \partial_{\kappa_1,\dots,\kappa_s}\lr\det\lp Z + \sum_{i=1}^s\kappa_i W_i\rp \rs_{\kappa_i = 0}
\eeq
We find that for one and two excitations
\begin{align}
\label{eq:oneex}
\Ocal(W_1) &= \det(Z)\Tr(Z^{-1}W_1) \\
\label{eq:twoex}
\Ocal(W_1,W_2) &= \det(Z)(\Tr(Z^{-1}W_1)\Tr(Z^{-1}W_2) - \Tr(Z^{-1}W_1Z^{-1}W_2))
\end{align}
We see that the operator corresponding to a maximal giant can be separated from the excitations.
The boundary conditions for opens string in the gauge theory are expressed as poles in specially crafted multi-trace operators.

These giant graviton states have definite angular momenta, but are delocalized in position space on the $S^5$.
To establish a connection to the classical string theory, we want operators that are localized in position space and have indefinite angular momentum.
This is accomplished by shifting the matrix $Z$ by a complex collective coordinate $\alpha$, that is, we take $\det(Z) \to \det(Z - \alpha)$ \cite{Berenstein:2013md}.
The coordinate $\alpha$ lives in a complex unit disk subspace of the $S^5$ normalized by the radius.
Expanding the operator $\det(Z-\alpha)$ in $\alpha$, we see that it is actually a coherent state of giant gravitons with different angular momenta.
Thus it is classical in this sense as well.

Given a set of giant gravitons and their collective coordinates $\alpha_i$, we can write down an approximate generating function that allows us to stretch open strings from one giant to another.
Let $W_{mn}$ be a string excitation connecting from giant graviton $m$ to giant graviton $n$.
In light of \eqref{eq:genfunc} we have
\begin{align}
\label{eq:mgiants}
M &=
\begin{pmatrix}
Z - \alpha_1 & \kappa_{12}W_{12} & \kappa_{13}W_{13} & \cdots \\
\kappa_{21}W_{21} & Z - \alpha_2 & \kappa_{23}W_{23} & \cdots \\
\kappa_{31}W_{31} & \kappa_{32}W_{32} & Z - \alpha_3 & \cdots \\
\vdots & \vdots & \vdots & \ddots
\end{pmatrix} \\
\label{eq:genfuncgen}
\Ocal(\{W_{mn}\}) &= \partial_{\{\kappa_{mn}\}}\lr \det(M)\rs_{\kappa_{mn} = 0}
\end{align}
The operator $\Ocal(\{W_{mn}\})$ will be non-vanishing only if the set $\{\kappa_{mn}\}$ form a cycle over some sequence of D-branes.
That is, given a sequence of the D-branes $\{i_k\}_{k=1}^s$ the $\kappa$'s must be able to be organized into the sequence 
\beq
\kappa_{i_1i_2}\kappa_{i_2i_3}\dots\kappa_{i_{k-1}i_k}\kappa_{i_{k}i_{k+1}}\dots\kappa_{i_{s-1}i_{s}}\kappa_{i_{s}i_1}
\eeq
To see this, let us consider the case of three giant gravitons with string excitations as present in \eqref{eq:mgiants}.
Let $\partial_i$ denote a derivative with respect to one of the $\kappa_{jk}$ which for now we leave unspecified.
Taking three derivatives of $\det(M)$ we have
\begin{align}
\partial_3\partial_2\partial_1\det(M) &= \det(M)\lp\Tr(M^{-1}\partial_3M)\Tr(M^{-1}\partial_2M)\Tr(M^{-1}\partial_1M) \rr \nonumber \\
&\qquad - \Tr(M^{-1}\partial_3M)\Tr(M^{-1}\partial_2M M^{-1}\partial_1M) \nonumber \\
&\qquad - \Tr(M^{-1}\partial_1M)\Tr(M^{-1}\partial_3M M^{-1}\partial_2M) \nonumber \\
&\qquad - \Tr(M^{-1}\partial_2M)\Tr(M^{-1}\partial_1M M^{-1}\partial_3M) \nonumber \\
&\qquad + \Tr(M^{-1}\partial_3MM^{-1}\partial_2M M^{-1}\partial_1M) \nonumber \\
\label{eq:examplederiv}
&\qquad + \lr\Tr(M^{-1}\partial_3MM^{-1}\partial_1M M^{-1}\partial_2M)\rp
\end{align}
After setting $\kappa_{jk} = 0$, $M^{-1}$ is diagonal with blocks $(Z - \alpha_i)^{-1}$.
If $\partial_i = \partial_{\kappa_{jk}}$, then $\partial_i M$ will contain a single string excitation $W_{jk}$ at row $j$ and column $k$.
Since $M^{-1}|_{\kappa_{mn}=0}$ is diagonal, $M^{-1}\partial_i M|_{\kappa_{mn}=0}$ will also only contain a single entry at row $j$ and column $k$.
Consecutive products of $M^{-1}\partial_i M|_{\kappa_{mn}=0}$ will be non-vanishing if the column of the string excitation in the left prodand matches the row of the string excitation of the right prodand.
The trace implies that this must occur between the first and last occurances of $M^{-1}\partial_i M|_{\kappa_{mn}=0}$.
Therefore any collection of desired string excitations must form a cycle over the connecting branes.
An example with three giant gravitons, with string stretching from branes $1\to 2\to 3\to 1$ is given by
\begin{align}
&\hspace{-2cm} \det(Z - \alpha_1)\det(Z - \alpha_2)\det(Z - \alpha_3) \nonumber \\
&\quad \times \Tr\lp(Z -\alpha_1)^{-1}W_{12}(Z - \alpha_2)^{-1}W_{23}(Z - \alpha_3)^{-1}W_{31}\rp
\end{align}

We can generalize \eqref{eq:genfuncgen} to include self-excitations $W_{mm}$ and multiple excitations between branes $W_{mn}^{(i)}$.
Regardless, the cyclic property of the $\kappa$'s remains the same.
Thus the number of string excitations leaving a D-brane must be equal to the number of string excitations entering a D-brane.
A Gauss' law for the number of open strings attached to a D-brane has emerged and gives rise to a $U(1)$ gauge symmetry \cite{Berenstein:2013md}.
The Gauss' law tells us that the total number of excess strings at each brane in the configuration has to be zero.
The Gauss' law disallows monopoles, but allows for dipole configurations.
These have been studied in \cite{Sadri:2003mx} for a $U(1)$ gauge theory on D-branes with worldvolume $\mathbb{R}\times S^3$ in a plane wave background.

It is tempting to interpret the product of determinants as a coherent state of bound giant graviton states, however, the products of determinants do not have the proper orthogonality properties to be interpreted this way.
Thus the generating function \eqref{eq:mgiants} is really only an approximation to the physical situation we are interested in.
Introducing the generating function is a simple, intuitive way of seeing the emergence of a Gauss' law for open strings.

Correctly dealing with multiple giant gravitons was originally worked out in \cite{Corley:2001zk}.
The subdeterminant operators \eqref{eq:subdet} are Schur polynomials in the matrix $Z$ corresponding to a Young diagram with one column and $k$ boxes.
The Young diagram correpsonds to a representation of the symmetric group and  dictates how to contract the indices over many occurrences of the matrix $Z$.
The representation can be arbitrary and its interpretation as point-like gravitons, strings, giant gravitons, and dual giant gravitons depends on the number and length of the columns and rows in the correpsonding Young diagrams.
In particular, long columns are thought of as the giant gravitons living in $S^5$ while long rows are thought of as the dual giant gravitons living in $AdS_5$.
To be concrete, if $R$ is a representation of the symmetric group $S_n$, then the corresponding graviton operator is given by
\beq
\chi_R(Z) = \sum_{\sigma\in S_n} \chi_R(\sigma) Z_{i_{\sigma(1)}}^{i_1}Z_{i_{\sigma(2)}}^{i_2}\cdots Z_{i_{\sigma(n)}}^{i_n}
\eeq
where $\chi_R(\sigma)$ is the character of the group element $\sigma$ in the representation $R$ and all indices $i_j$ (which range from $1$ to $N$) are summed over.
Different representations are orthogonal with respect to free field theory contractions
\beq
\langle \chi_R(\bar{Z})\chi_S(Z)\rangle = \delta_{RS}f_R
\eeq
with $f_R$ the weight of the Young diagram corresponding to the representation $R$.

String excitations $W_i$ are added to the giants by introducing restricted characters for representations of the symmetric group.
These restrictions fix where the gauge indices are contracted to the rest of the giant graviton operator.
The details are beyond the scope of this work and we defer the reader to the review articles \cite{deMelloKoch:2007uu,deMelloKoch:2007uv,Bekker:2007ea}.
Since these giants are spherical in nature and thus topologically compact, we expect a Gauss' law constraint for the number of open strings.
Although this has not been proven, there is convincing evidence \cite{Balasubramanian:2004nb} that this construction respects the Gauss' law constraint.

Instead of delocalizing bound states of multiple giants expressed as Schur polynomials, we simplify the problem of strings stretched between giant gravitons by considering supersymmetric orbifolds of the $\Ncal = 4$ SYM theory \cite{Berenstein:2013md}.
Multiple giants with collective coordinates $\alpha_i$ can then be expressed as a product of determinant operators
\beq
\Ocal(\alpha_1,\dots,\alpha_k) = \det(Z_1 - \alpha_1)\cdots \det(Z_k - \alpha_k)
\eeq
with each $Z_i$ belonging to a different sector of the orbifolded theory.
We stretch strings between the giant given the schematic of \eqref{eq:oneex} and \eqref{eq:twoex}.
That is, a pole in the field $Z_i - \alpha_i$ is introduced when we want to connect strings beginning and ending on the giant graviton $\alpha_i$.
For example, the operator for three giant gravitons with strings stretching from branes $1\to 2\to 3\to 1$ is given by
\begin{align}
&\det(Z_1 - \alpha_1)\det(Z_2 - \alpha_2)\det(Z_3 - \alpha_3) \times \nonumber \\
&\quad \hphantom{\det(Z_1 - \alpha_1)} \Tr\lp(Z_1 -\alpha_1)^{-1}W_{12}(Z_2 - \alpha_2)^{-1}W_{23}(Z_3 - \alpha_3)^{-1}W_{31}\rp
\end{align}

Up to this point we have chosen to leave the string excitations $W_i$ arbitrary.
We restrict the excitations $W_i$ to be composed only of the fields $Y$ and $Z$. Then the giant graviton operators together with their stringy excitations live in the $\sutwo$ sector of the gauge theory, where a great deal of calculations are accessible from both sides of the duality.


\section{Ground State of the Open String}
\label{sec:dilatation}

The operators of Section \ref{sec:openstrings} give a concrete handle by which to manage open strings directly in the gauge theory.
Of particular interest is the open string ground state and spectrum of excited states above it.
These can be acquired by diagonalizing the dilatation generator on this space of states.
The action of the dilatation generator in the $\sutwo$ sector is known to two-loops and in the planar limit up to five-loops.
In the planar limit, the dilatation generator acts on single trace operators.
Although the open string states are not single trace operators, the inclusion of the boundary conditions is simple as we will later discuss.

The dilatation generator in the $\sutwo$ sector can be written as an expansion in the 't Hooft coupling $\lambda = g_{\text{YM}}^2N$
\beq
\label{eq:dgen}
D = \sum_{k=0}^{\infty} \lp\frac{\lambda}{4\pi^2}\rp^k D_k
\eeq
where $k$ represents the number of loops.
The first three terms in this sequence are given by \cite{Beisert:2003tq}
\begin{align}
\label{eq:d0field}
D_0 &= \mathop{:}\Tr(Z\Zch)\mathop{:} + \mathop{:}\Tr(Y\Ych)\mathop{:} \\
\label{eq:d1field}
D_1 &= -\frac{1}{2N}\mathop{:}\Tr([Y,Z][\Ych,\Zch])\mathop{:} \\
\label{eq:d2field}
D_2 &= -\frac{1}{8N^2}[\mathop{:}\Tr([[Y,Z],\Ych][[\Ych,\Zch],Y])\mathop{:} + \mathop{:}\Tr([[Y,Z],\Zch][[\Ych,\Zch],Z])\mathop{:} \nonumber \\
&\phantom{= -\frac{1}{8N^2}[} + \mathop{:}\Tr([[Y,Z],T^a][[\Ych,\Zch],T^a])\mathop{:}]
\end{align}
where $T^a$ are generators of $U(N)$ ($SU(N)$), the fields are expanded in these generators $Y = Y^aT^a$, and the checked operators are derivatives with respect to that field, $\Zch_A^B Z_C^D = (T^a)_A^B (T^b)_C^D \Zch^a Z^b = (T^a)_A^B (T^b)_C^D \delta^{ab} = \delta_A^B\delta_C^D$.
The normal ordering indicates that derivatives do not act on other fields inside the normal ordering.
The states include single and multi-trace operators of the fields $Y$ and $Z$.
In the planar limit, the action of the dilatation operator simplifies by acting on individual single trace factors.

The space of states can then be represented as an $\sutwo$ spin chain with each type of field representing either spin up or down.
The one-loop dilatation generator was recognized as the Hamiltonian of the Heisenberg spin chain and helped kick off the integrability program in the context of the AdS / CFT correspondence.
Local interactions can be represented as permutations of nearest neighbors, next-to-nearest neighbors, etc.
The dilatation generator can also be represented in terms of the so-called chiral functions \cite{Fiamberti:2007rj}.
They are obtained by shifting the permutation operators by the identity.
They represent the chiral structure of the underlying Feynman super-graphs that generate them \cite{Sieg:2010tz}.

The dilatation operator in the $\sutwo$ sector does not change the length of the spin chain; the total number of fields is conserved.
The Cartan generator of the $\sutwo$ yields the difference between the number of $Y$ and $Z$ fields.
It commutes with the dilatation generator and hence is also conserved.
Thus the individual number of $Y$ and $Z$ fields is unchanged.
We can choose $Z$ to be a highest weight state of the $\mathfrak{so}(6)$ R-symmetry which is positively charged under one of the angular momenta $\Jhh_1$.
In the $\mathfrak{su}(2)$ sector one can represent $\Jhh_1 = \Tr(Z\Zch)$.
We can do something similar with $Y$ by choosing an orthogonal momentum $\Jhh_2 = \Tr(Y\Ych)$.
These are dual to the two orthogonal momenta on the $S^3$ in the $\mathbb{R}_t\times S^3$ subspace of $AdS_5\times S^5$.
The tree level dilatation operator is then given by $D_0 = \Jhh_1 + \Jhh_2$.

Another basis has been constructed to represent the space of states in the planar $\sutwo$ sector.
To construct them we fix the number of $Y$ fields in a single trace.
Any number of $Z$ fields can be placed between the $Y$.
That is, the $Y$ separate sites at which one can place $Z$ fields.
Thus a state is labeled by the number of $Y$'s and the number of $Z$'s between each $Y$.
The cyclic nature of the trace means that we should cyclically identify the states.
For example 
\beq
\label{eq:closedcuntzchain}
\Tr(ZYZZZYZZY) \to \ket{3,2,1} \equiv \ket{2,1,3} \equiv \ket{1,3,2}
\eeq
A general state is represented by some set of occupation numbers $\{n_i\}_{i=1}^k$, where $k$ is the number of $Y$ fields.
The zero occupation number state is just $\Tr(Y^k)$, which we denote as $\ket{0}_k$.
To represent insertions of the field $Z$ we introduce a set of Cuntz oscillators \cite{cuntz1977} $a_i$, $a_i^\dag$ which satisfy the commutation relations $[a_i, a_j^\dag] = \delta_{ij}P_{0i}$ with $P_{0i}$ the projection onto the zero number occupation state at site $i$.
We call states of the form \eqref{eq:closedcuntzchain} closed Cuntz oscillator chains.
The dilatation generator computed in this basis \cite{Berenstein:2013eya} is given by
\begin{align}
\label{eq:d0cuntz}
D_0 & \rightarrow  k + \sum_{i=1}^k n_i \\
\label{eq:d1cuntz}
D_1 & \rightarrow H_{\text{closed},0} = \frac{1}{2}\sum_{i=1}^k (a_{i+1}^\dag - a_i^\dag)(a_{i+1} - a_i) \\
\label{eq:d2cuntz}
D_2 & \rightarrow H_{\text{closed},1} = -\frac{1}{8}\sum_{i=1}^k (a_{i+1}^\dag - a_i^\dag)^2(a_{i+1} - a_i)^2 \nonumber \\
&\hphantom{\rightarrow H_{\text{closed},1} = -\frac{1}{8}\sum_{i=1}^k} + (a_{i+1}^\dag - 2a_{i}^\dag + a_{i-1}^\dag)P_{0i}(a_{i+1} - 2a_{i} + a_{i-1})
\end{align}
where the $a_i$ are identified cyclically, $a_{i+k} \equiv a_{i}$.
The dilatation generator becomes a Hamiltonian like energy operator on the Cuntz oscillator chains.
We will thus refer to the dilatation generator in the Cuntz basis as the Cuntz Hamiltonian and its eigenvalue as the energy.
The closed Cuntz chains are isomorphic to the space of single trace operators.
The anomalous dimensions computed using $\eqref{eq:d0cuntz}-\eqref{eq:d2cuntz}$ are equal to those computed using the dilatation generator in the basis of permutation operators or chiral functions.

The Cuntz oscillator basis can be used to represent open strings as well.
Consider an open string state with strings stretched between two giant gravitons with collective coordinates $\alpha_1$, $\alpha_2$.
As a reminder, for multiple giants it is simpler to work in the supersymmetric $\mathbb{Z}_n$ orbifold of $\Ncal=4$ SYM.
For $n=2$, this corresponds to a $U(N)\times U(N)$ quiver theory with $\Ncal = 2$ SUSY in four dimensions.
The chiral superpartners of the vector fields will be called $Z_1,~Z_2$, while the matter hypermultiplets between the two gauge groups will be made of $X,~Y$ chiral fields. 
The corresponding state will be given by
\beq
\label{eq:openspinchainorbifold}
\det(Z_1 - \alpha_1) \det(Z_2 - \alpha_2)\Tr\lp\frac{1}{Z_1 - \alpha_1}W_{12}\frac{1}{Z_2 - \alpha_2} W_{21}\rp
\end{equation}
The string excitation $W_{21}$ is needed to satisfy the Gauss' law constraint.
The dilatation generator acts on single trace factors in a multi-trace operator in the planar limit.
The analogy for open strings is that the dilatation generator will not act on $W_{12}$ and $W_{21}$ simultaneously without splitting the trace in the planar limit.
Thus we can focus our attention on individual strings stretched between giant gravitons instead of the entire cycle, say $W_{12}$.
The string excitation $W_{12}$ is now given in this orbifolded case by
\beq
W_{12} = Y_{12} Z_2^{n_1} Y_{21} Z_1^{n_2} Y_{12} Z_2^{n_3} \cdots Z_1^{n_k} Y_{12}
\eeq
The labels $Y_{12}$ indicate that the $Y$ is a bifundamental in the $(N_1, \bar N_2)$ representation of the $U(N_1)\times U(N_2)$ orbifold group (with $N_1=N_2=N$ numerically), whereas the $Y_{21}$ is in the $(\bar N_1, N_2)$ representation.
Thus an open string stretched from $\alpha_1$ to $\alpha_2$ can also be represented by a Cuntz oscillator chain with the $Y$ separating the individual sites.
We take the open string states to be of the form $\ket{\{n_i\}_{i=1}^k}$ without cyclic identification of the sequence.
One should keep in mind that for these states, even though we have only $k$ sites, there are $k+1$ $Y$ fields.
This is to separate the giant graviton pole from other $Z$ fields.

The planar dilatation generator can not be directly applied to the open string states.
It acts on the string excitation $W_{12}$ as it would on a single trace operator as the boundary is not detectable in the middle of the string.
One has to additionally include the effects of the poles in the trace and the determinants.
This can only be done by using the dilatation generator in the field basis $\eqref{eq:d0field}-\eqref{eq:d2field}$.
The effects of the boundary have been computed at one and two-loops \cite{Berenstein:2013eya}.
States of the form \eqref{eq:openspinchainorbifold} were used to simplify the calculation.
We are forced to have an even number of sites because $Z_1$ and $Z_2$ alternate between each other in the chain.
However, this does not affect the boundary conditions on the chain that we want to derive and we will ignore this condition.

The result of the boundary calculation is to modify the operators $\eqref{eq:d0cuntz} - \eqref{eq:d1cuntz}$ by breaking the cyclicity of the Cuntz chain and add auxiliary boundary sites whose Cuntz operators are replaced by ordinary c-numbers.
The c-numbers are the collective coordinates of the giant gravitons scaled by a factor of $\sqrt{N}$ denoted by $\xi = \alpha^*_1 / \sqrt{N}$ and $\xitt = \alpha_2^* / \sqrt{N}$.
Explicitly one has
\begin{align}
\label{eq:d1cuntzopen}
D_1 \rightarrow H_{\text{open},0} &= \frac{1}{2}\sum_{i=0}^k (a_{i+1}^\dag - a_i^\dag)(a_{i+1} - a_i) \\
\label{eq:d2cuntzopen}
D_2 \rightarrow H_{\text{open},1} &= -\frac{1}{8}\sum_{i=0}^k (a_{i+1}^\dag - a_i^\dag)^2(a_{i+1} - a_i)^2 \nonumber \\
&\hphantom{= -\frac{1}{8}\sum_{i=0}^k} -\frac{1}{8}\sum_{i=1}^{k} (a_{i+1}^\dag - 2a_{i}^\dag + a_{i-1}^\dag)P_{0i}(a_{i+1} - 2a_{i} + a_{i-1})
\end{align}
where
\beq
\label{eq:bidentify}
a_0 = \xi, \quad a_0^\dag = \xi^*, \quad a_{k+1} = \xitt, \quad a_{k+1}^\dag = \xitt^*
\eeq
A comment on the tree level dilatation generator is in order.
The operator $D_0 = \Jhh_1 + \Jhh_2$ just counts the number of $Y$ and $Z$ fields.
Although $\Jhh_2 = k + 1$ is well defined, the determinants contribute order $N$ number of $Z$ fields.
Thus $\Jhh_1$, and hence its anomalous dimension, diverges in the planar limit.
We should instead consider the combination $D - \Jhh_1$ for open strings.
In particular we translate the tree level dilatation generator to
\beq
\label{eq:d0cuntzopen}
(D_0 - \Jhh_1) \rightarrow k+1
\eeq

The ground state of the one-loop dilatation generator \eqref{eq:d1cuntzopen} has been computed using coherent states of the Cuntz oscillators.
We introduce a set of collective coordinates for the sites $\{z_i\}_{i=1}^k$ such that $a_i\ket{\vec{z}\,} = z_i\ket{\vec{z}\,}$.
The string collective coordinates can be interpreted as bits of string that form a non-trivial quantum gas.
The ground state to this order in perturbation theory, denoted by $\kgs \equiv \ket{\vec{z}^{\,0}}$, is given by a set of collective coordinates that linearly interpolate between the two giant gravitons' scaled collective coordinates
\beq
\label{eq:groundstate}
z_i^0 = \frac{1}{k+1}((\xitt - \xi)i + (k + 1)\xi)
\eeq
The ground state $\kgs$ is an eigenstate of the dilatation generator at zero and one-loop, but this is no longer the case at two-loops.
The energy gets corrected at two-loops through standard perturbation theory, that is, it is just an expectation value.
Let $\ket{\Omega}$ denote the full ground state and $H_{\text{open}}$ the translation of the planar limit of $D - \Jhh_1$ to the open Cuntz oscillator chain basis.
The energy of the full ground state given by $H_{\text{open}}\ket{\Omega} = E_0\ket{\Omega}$ has an expansion in the 't Hooft coupling
\beq
\label{eq:e0gaugeapprox}
E_0 = (k+1) + \frac{1}{2}\lp\frac{\lambda}{4\pi^2}\rp\frac{|\xi - \xitt|^2}{k+1} - \frac{1}{8}\lp\frac{\lambda}{4\pi^2}\rp^2\frac{|\xi - \xitt|^4}{(k+1)^3} + \Ocal(\lambda^3)
\eeq
The expansion of $E_0$ is consistent with the all-loop relativistic dispersion relation
\beq
\label{eq:e0gauge}
E_0 = \sqrt{(k+1)^2 + \frac{\lambda}{4\pi^2}|\xi - \xitt|^2}
\eeq
as was conjectured in \cite{Berenstein:2013eya}.
This relation is relativistic roughly because it takes the form $E = \sqrt{m^2 + p^2}$ by appropriately identifying the mass, $m$, and momentum, $p$.
In flat space, separating D-branes generates massive $W$ bosons whose mass is proportional to the separation between the two branes.
Thus we should identify the mass of the open string with the separation between the giant gravitons, $m \propto |\xi - \xitt|$.
The momentum is identified with the angular momentum, $p = k+1$, of the string propagating in $S^3\subset S^5$.
With these identifications, we see that the energy \eqref{eq:e0gaugeapprox} is an expansion in the mass.

A more concrete way to see \eqref{eq:e0gauge} as relativistic is to show that it comes from the local gauge theory on the world volume of the giant gravitons.
Consider $\Ncal = 4$ SYM with a $U(2)$ gauge symmetry on $\mathbb{R}\times S^3$ living on the two gravitons with zero separation.
This $S^3$ is the world volume of the giants and not the $S^3$ of the dual gauge theory in the AdS / CFT correspondence.
The (local) kinetic operator for the scalar fields on the giant gravitons is
\beq
-\partial_t^2 + \nabla^2_{S^3} - \frac{1}{\rgg^2}
\eeq
with $\rgg$ the radius of the spherical giants and $\rgg^{-2}$ the coupling to the curvature of the sphere necessary for conformal invariance.
Performing a Fourier transform, we have the eigenvalue equation
\beq
0 = E^2 - \frac{\ell(\ell + 2)}{\rgg^2} - \frac{1}{\rgg^2} = E^2 - \frac{(\ell + 1)^2}{\rgg^2}
\eeq
with $\ell$ the quantum number of the corresponding harmonic on the sphere.
If we separate the giant gravitons, the gauge fields acquire a constant background value breaking the gauge symmetry to $U(1)\times U(1)$ and the scalars acquire a mass proportional to the difference between the background values, that is, the separation between the branes.
The dispersion relation for the strings becomes
\beq
E^2 = \frac{(\ell + 1)^2}{\rgg^2} + m^2
\eeq
Normalizing the radius of the giant gravitons and substituting in the correct constant of proportionality for the mass we obtain \eqref{eq:e0gauge} showing that it is local and hence relativistic.

The strings dual to these open string ground states was given in \cite{Berenstein:2014isa}.
They are obtained by truncating the solitonic strings on $AdS_5\times S^5$ propagating in the $\mathbb{R}_t\times S^3$ subspace.
The $S^3$ is parametrized by two complex coordinates $Z_1$, $Z_2$, each living in the complex unit disk with the restriction that $|Z_1|^2 + |Z_2|^2 = 1$.
The solitonic strings are characterized by straight lines running between the boundary of the disk in which $Z_1$ lives.
The open string solutions are obtained by restricting the extent of the lines to end on the collective coordinates of giant gravitons instead of the edge of the disk.
This description matches precisely with the collective coordinates $z_i^0$ of the open string ground state produced above.
Let $E$ be the energy of the string, and $J_1$ and $J_2$ denote the orthogonal angular momenta on the $S^3$ of the string.
These conserved quantities were shown to be related to each other by
\beq
\label{eq:e0string}
E - J_1 = \sqrt{J_2^2 + \frac{\lambda}{4\pi^2}|\xi - \xitt|^2}
\eeq
The $Y$ fields in the gauge theory contribute to $\Jhh_2$ and so we should identify $k+1$ with the string theory $J_2$.
They add single units of angular momentum to the dual string in the $S^3$.
We additionally identify $D - \Jhh_1$ in the gauge theory with $E - J_1$ because the dilatation operator on $\bbr^4$ becomes the Hamiltonian on $\bbr\times S^3$ after radial quantization and it is this energy that is dual to the $AdS_5$ energy in global coordinates.
The relation \eqref{eq:e0string} and the configuration of the open string solution gives supporting evidence that these open strings are indeed dual to ground states of the Cuntz oscillator Hamiltonian characterized by the collective coordinates \eqref{eq:groundstate}.

What we would like to do is confirm \eqref{eq:e0gauge} to second order in perturbation theory in the context of the gauge theory.
At this order in perturbation theory, the full spectrum of eigenvalues of the Cuntz Hamiltonian makes a contribution.
The boundary conditions take us away from maximal giants, so we do not expect any integrability to hold.
As such it is unexpected that we can compute the spectrum analytically.
Despite this limitation, we can find a basis in which to compute the first order correction to the ground state and the second order correction to the energy analytically.
In order to go the next order in perturbation theory, we will need to translate the three-loop dilatation operator into the Cuntz language.


\section{The Cuntz Hamiltonian at Three Loops}
\label{sec:threeloops}

We start with the planar dilatation generator up to three-loops in the basis of chiral functions given by 
\cite{Sieg:2010tz}
\begin{align}
\label{eq:d1chiral}
D_1 &= -\frac{1}{2}\chi(1) \\
\label{eq:d2chiral}
D_2 &= -\frac{1}{8}[(\chi(1,2) + \chi(2,1)) - 2\chi(1)] \\
\label{eq:d3chiral}
D_3 &= -\frac{1}{16}[(\chi(1,2,3) + \chi(3,2,1)) + \chi(1,3) - 4(\chi(1,2) + \chi(2,1)) + 6\chi(1)]
\end{align}
We want to translate this into the Cuntz oscillator picture.
This was carried out at one and two-loops using the dilatation generator in the field basis.
It is much simpler to do this in the basis of chiral functions, but the calculation is tedious and so we refer the reader to the Appendix \ref{sec:trans} for further details.
The result for the closed Cuntz chain is
\begin{align}
D_3 \rightarrow H_{\text{closed},2} &= \frac{1}{16}\sum_{i=1}^k (a_{i+1}^\dag - a_{i}^\dag)^3(a_{i+1} - a_{i})^3 + v_a^{i\dag} M_{ab}P_{0i}v_b^i \nonumber \\
\label{eq:d3cuntz}
&\qquad + (a_{i+2}^\dag - 3a_{i+1}^\dag + 3a_{i}^\dag - a_{i-1}^\dag)P_{0i+1}P_{0i}(a_{i+2} - 3a_{i+1} + 3a_{i} - a_{i-1})
\end{align}
where $v_a$ is a vector of operators
\beq
v_a^i = (a_{i+1}a_{i+1},~a_{i+1}a_{i},~a_{i+1}a_{i-1},~a_{i}a_{i},~a_{i}a_{i-1},~a_{i-1}a_{i-1})
\eeq
and $M_{ab}$ is the matrix
\beq
\label{eq:stmat}
M_{ab} = %
\begin{pmatrix}
 4 & -8 &  2 &  2 &  0 &  0 \\
-8 & 15 & -3 & -3 & -1 &  0 \\
 2 & -3 &  1 &  1 & -3 &  2 \\
 2 & -3 &  1 &  1 & -3 &  2 \\
 0 & -1 & -3 & -3 & 15 & -8 \\
 0 &  0 &  2 &  2 & -8 &  4
\end{pmatrix}
\eeq
The symmetry of $M_{ab}$ is necessary for the Cuntz Hamiltonian to be Hermitian.

Before commenting on each term in \eqref{eq:d3cuntz}, we would like to modify it for the open string.
As at lower loop orders, the effects of the boundary need to be computed with the dilatation operator in the field basis.
However, it is not enough to translate the planar dilatation generation in terms the basis of chiral functions into planar contributions in the field basis.
In \cite{Berenstein:2013eya} it was shown that non-planar contributions to the dilatation generator have finite contribution in the planar limit when acting on the determinants because their scaling grows linearly with $N$.
The full three-loop non-planar dilatation generator in the $\sutwo$ sector is not known at this time, although one should be able to compute it by extending the analysis of \cite{Sieg:2010tz} to non-planar graphs.

At one and two-loops the boundary had the simple effect of breaking the period of the closed Cuntz chain and adding two boundary `sites' that were ordinary c-numbers.
Understanding the action of the dilatation operator at one and two-loops on delocalized giants $\det(Z - \alpha)$ %
\cite{Berenstein:2013md,Berenstein:2013eya} and expanding in the collective coordinate $\alpha$, we see that the subdeterminants can be realized as a truncated representation of the harmonic oscillator algebra.
Denote the states by $\ket{n} = \det_n(Z)$ and operators $b$ and $b^\dag$ that act as $b\ket{n} = N^{-1/2}\sqrt{n}\ket{n}$, $b^\dag\ket{n} = N^{-1/2}\sqrt{n+1}\ket{n}$.
The delocalized giant graviton operators can be realized as a coherent state of this truncated algebra $\ket{\alpha} = \det(Z - \alpha)$ which satisfy $b\ket{\alpha} = \frac{\alpha}{\sqrt{N}}\ket{\alpha}$ \cite{Berenstein:2014zxa}.
Then the one-loop Cuntz Hamiltonian should be written as
\beq
\label{eq:d1cuntzopent}
\sim (b_2^\dag - a_{k}^\dag)(b_2 - a_{k}) + \sum_{i=1}^{k-1}(a_{i+1}^\dag - a_{i}^\dag)(a_{i+1} - a_{i}) + (a_1^\dag - b_1^\dag)(a_1 - b_1)
\eeq
where the subscript on $b$ denotes which giant graviton the operator acts on.
The one-loop open Cuntz chain Hamiltonian \eqref{eq:d1cuntzopen} is then realized as \eqref{eq:d1cuntzopent} in a background of coherent state giant gravitons.
This approximation is valid because because we are taking the planar limit where the D-branes are rigid objects and the open strings do not backreact.
We will assume that the giant gravitons can be inserted into the Cuntz chain as ordinary harmonic oscillators at three-loops as well.
Realizing we are taking the planar limit then allows us to replace the operators with ordinary c-numbers.
The open Cuntz Hamiltonian at three-loops is then given by
\begin{align}
D_3 \rightarrow H_{\text{open},2} &= \frac{1}{16}\sum_{i=0}^k (a_{i+1}^\dag - a_{i}^\dag)^3(a_{i+1} - a_{i})^3 + \frac{1}{16}\sum_{i=1}^{k} v_a^{i\dag} M_{ab}P_{0i}v_b^i \nonumber \\
\label{eq:d3cuntzopen}
&\qquad + \frac{1}{16}\sum_{i=1}^{k-1} (a_{i+2}^\dag - 3a_{i+1}^\dag + 3a_{i}^\dag - a_{i-1}^\dag)P_{0i+1}P_{0i}(a_{i+2} - 3a_{i+1} + 3a_{i} - a_{i-1})
\end{align}
with the identification \eqref{eq:bidentify}.

A few comments are in order regarding the three terms of \eqref{eq:d3cuntz}, and similarly \eqref{eq:d3cuntzopen}.
The first term of \eqref{eq:d3cuntz} is the magnitude squared of the cube of a first derivative finite difference between two sites.
Similar terms appear in \eqref{eq:d1cuntz} and \eqref{eq:d2cuntz} except that they are raised to the first and second powers respectively.
One might then expect that at $\ell$-loop order the open Cuntz Hamiltonian should receive a contribution
\beq
c_\ell\sum_{i=0}^k (a_{i+1}^\dag - a_{i}^\dag)^\ell(a_{i+1} - a_{i})^\ell
\eeq
with $c_\ell$ some constant.
Given the numerical expansion \eqref{eq:dgen}, the expectation value of these operators of the open string ground state $\kgs$ yield the naive contributions to the relativistic dispersion relation \eqref{eq:e0gauge} if $c_\ell$ takes the value
\beq
c_\ell = \frac{1}{2\ell - 1}\frac{(2\ell)!}{(\ell!)^2 2^{2\ell}}(-1)^{\ell+1}
\eeq
These first derivative terms come from chiral functions of the form $\chi(1,2,\dots,\ell) + \chi(\ell,\dots,2,1)$.
Their coefficient should be $c_\ell(-1)^\ell$.
Looking at the relevant permutation structures of the dilatation operator in \cite{Beisert:2004ry}, we find agreement up to five-loops %
\footnote{See table 6.6 \cite{Beisert:2004ry} for the four and five-loop operators.
Instead of looking at the chiral functions $\chi(1,2,\dots,\ell) + \chi(\ell,\dots,2,1)$, it suffices to look at the permutation structures $\{1,2,\dots,\ell\} + \{\ell,\dots,2,1\}$.
One should multiply by $2^\ell$ due to the conventions for the coupling constant and expansion of the dilatation generator.}.

The last term of \eqref{eq:d3cuntz} is the magnitude squared of a third derivative finite difference operator.
There are two projection operators necessary for $SU(2)$ invariance.
They take into account consecutive sites with low occupation number.
These are operators where the $Y$ fields are separated by at most two $Z$.
The last term is also symmetric about the four sites on which it operates.
Switching $i+2 \leftrightarrow i-1$ and $i+1\leftrightarrow i$ is a symmetry of this term.
An analogous term also appears at two-loops, except that it is a second order finite difference, not a third, and there is one projector, not two.
At two-loops it is still symmetric under reflection of the three sites involved.
At $\ell$-loops we expect a similar term to occur; the magnitude squared of an $\ell$-order finite difference with $\ell-2$ projectors in the middle
\beq
\sim\sum_{i=1}^k \lp\sum_{m=0}^{\ell} {\ell\choose m} a_{i+m-\lfloor\ell/2\rfloor}^\dag\rp \lp\prod_{n=1}^{\ell-1} P_{0,i+n-\lfloor\ell/2\rfloor}\rp\lp\sum_{p=0}^{\ell} {\ell\choose p} a_{i+p-\lfloor\ell/2\rfloor}\rp
\eeq
These terms annihilate the open string ground state $\kgs$ and do not contribute to expectation values.

The second term of \eqref{eq:d3cuntz} is mysterious and has no analog at a lower order.
Despite numerous attempts by the author, a simple finite difference form of the operators quartic in the Cuntz oscillators could not be found.
Given the nature of the other terms, one would expect it to be a mixture of first and second order finite differences.
However, its expectation value against the open string ground state is non-vanishing, $\bgs(v_a^i)^\dag M_{ab}P_{0i} (v_b^i)\kgs \neq 0$.
Thus it can not be rewritten so that every term contains a second or higher order finite difference term.
The non-vanishing of the expectation value implies that the ground state {\it must} get corrected at higher orders in perturbation theory.
The second term, however, does have the same reflection symmetry as the others.
The exchange $i+1\leftrightarrow i-1$ is a site-by-site symmetry.
This can be seen by the fact that the matrix \eqref{eq:stmat} remains unchanged after flipping it over the horizontal and vertical axes and then swapping the middle two columns and rows.
The final swap is necessary because the exchange $i+1\leftrightarrow i-1$ does not swap $a_{i+1}a_{i-1}$ and $a_{i}a_{i}$ in $v_a^i$.

The reflection symmetry is a consequence of the existence of a parity symmetry at the planar level.
Thus we expect this to be a local symmetry of the Cuntz Hamiltonian to all orders in perturbation theory.
That is, the Cuntz Hamiltonian should have a normal ordered form such that every term exhibits the reflection symmetry at the site level.

Another universal property of all the terms in the Cuntz Hamiltonian at all loop levels here is that they contain the same number of Cuntz oscillators.
Recalling that $P_{0} = [a,a^\dag]$ is quadratic in the Cuntz oscillators, we see that all terms have $2\ell$ Cuntz oscillators at $\ell$-loop order.
Even though the different kinds of interactions have different range, they all have the same number of Cuntz oscillators when we fix the loop level.
This is different than the spin chain picture where the range of individual interactions can be different than the loop level (although the maximum range is bounded by the loop level).
We currently do not have an explanation for why the same number of Cuntz oscillators appear, however, we do predict that the Cuntz Hamiltonian can always be written in a form in which this is true at all loop orders.


\section{Excited States of the Open String}
\label{sec:excited}

The equations to second order in perturbation theory require sums over the eigenvalues of the Hamiltonian.
To compute the first order correction to the open string ground state $\kgs$, we need a handle on the excited states of the Cuntz Hamiltonian.
Even though the entire spectrum cannot be solved for, the tools developed here will help us solve for this correction.

We would like to construct a Fock space of states identifying the open string ground state as the vacuum.
Consider a single Cuntz excitation at site $i$, $a_i^\dag\kgs$.
It is not orthogonal to the ground state.
We substract off the projection and find that 
\beq
\bgs\lb(a_i^\dag - \zbb_i^0)\kgs\rb = 0
\eeq
We think of $(a_i^\dag - \zbb_i^0)\kgs$ as a one particle state.
The process for constructing an $n$-particle state is just the Gram-Schmidt process using the sequence of states $(a_i^\dag)^n\kgs$.
The result is that the $n$-particle states are given by
\beq
(a_i^\dag)^{n-1}(a_i^\dag - \zbb_i^0)\kgs
\eeq
They are mutually orthogonal, but not yet normalized.
We define the site operators
\beq
\label{eq:siteops}
A_i^{(n)} = 
\begin{cases}
1 & n = 0 \\
(a_i)^{n-1}(a_i - z_i^0) & n > 0
\end{cases}
\eeq
We include the $n=0$ case for convenience.
The operators \eqref{eq:siteops} annihilate the ground state for all $n>0$, $A_i^{(n)}\kgs = 0$.
We think of $A_i^{(n)\dag}$ as the creation operator for an $n$-particle state at site $i$.

One can show by induction that the site operators satisfy the commutation relations
\beq
\label{eq:sitecomm}
[A_i^{(m)}, A_j^{(n)\dag}] = \delta_{ij}\sum_{\ell=1}^{\min(m,n)} A_i^{(n-\ell)\dag}P_{0i}A_{i}^{(m-\ell)}
\eeq
Using \eqref{eq:sitecomm} we find that the states $A_i^{(n)\dag}\kgs$, $n > 0$, have norm squared
\beq
\ls\ls A_i^{(n)\dag}\kgs\rs\rs^2 = \bgs A_i^{(n)}A_i^{(n)\dag}\kgs = \bgs P_{0i}\kgs = 1 - |z_i^0|^2
\eeq
To fully normalize a state with particles created at multiple sites, one should divide by $(1 - |z_i^0|^2)^{1/2}$ for each $A_i^{(n)\dag}$ present.
Although for $m = n = 1$ we have
\beq
[A_i^{(1)}, A_j^{(1)\dag}] = \delta_{ij}P_{0i}
\eeq
the single excitation site operators do not form a Cuntz algebra.
This can be seen from the relation $a_ia_i^\dag = I$.
For the single excitation site operators we have
\beq
A_i^{(1)}A_i^{(1)\dag} = (I - |z_i^0|^2) - \zbb_i^0 A_i^{(1)} - z_i^0 A_i^{(1)\dag}
\eeq
As a consequence, if we choose to work in the basis of $n$-particle states instead of the occupation number basis, we can not simply make the replacement $a_i\rightarrow A_i^{(1)}$, $a_i^\dag\rightarrow A_i^{(1)\dag}$.
To translate from the Cuntz oscillators to the site operators, one must undo the transformation that takes $a_i^n$ to $A_i^{(n)}$.
For $n>0$ one has
\beq
\label{eq:transformback}
a_i^n = \sum_{\ell=0}^n (z_i^0)^\ell A_{i}^{(n-\ell)}
\eeq
The site operators have the property that they do not form an $\mathbb{N}$ homomorphism
\footnote{Thinking of $\mathbb{N}$ as a monoid, not a group.}.
That is, $A_i^{(m)}A_{i}^{(n)} \neq A_{i}^{(m + n)}$.
The composition rule for $m,~n > 0$ is $A_i^{(m)}A_{i}^{(n)} = A_{i}^{(m+n)} - zA_{i}^{(m+n-1)}$.
Thus the site operators do not build a true Fock space.
Regardless, these operators do generate orthogonal excitations above the open string ground state and we will continue to refer to them as particles.

Lastly we can extend the site operators to boundary of the open Cuntz chain by defining $A_{0}^{(1)} = b_1 - \xi$, $A_{k+1}^{(1)} = b_2 - \xitt$.
It only really makes sense to introduce the giant graviton collective coordinates when we are talking about coherent states.
Using the arguments of the previous section, we should then think of these operators in coherent state giant graviton backgrounds.
Then we have $A_{0}^{(1)} = A_{k+1}^{(1)} = 0$.

We can reexpress the one-loop Cuntz Hamiltonians \eqref{eq:d1cuntzopen} in terms of the site operators.
We define the difference between the string collective coordinates $\Delta z \equiv z_{i+1}^0 - z_{i}^0 = (\xitt - \xi) / (k+1)$.
We also define the energy of the ground state 
\beq
E_0^{(0)} = \frac{1}{2}\frac{|\xi - \xitt|^2}{k+1}
\eeq
The sum $\sum_{i=0}^k (A_{i+1} - A_{i})$ telescopes and vanishes.
One has
\begin{align}
H_{\text{open},0} &= \frac{1}{2}\sum_{i=0}^k (A_{i+1}^{(1)\dag} - A_{i}^{(1)\dag} + \Delta \zbb)(A_{i+1}^{(1)} - A_{i}^{(1)} + \Delta z) \nonumber \\
&= E_{0}^{(0)} + \frac{1}{2}\sum_{i=0}^k(A_{i+1}^{(1)\dag} - A_{i}^{(1)\dag})(A_{i+1}^{(1)} - A_{i}^{(1)})
\end{align}
Writing the Hamiltonian in this form directly isolates the ground state energy from the energy of the excited states.
That is, since $A_i^{(1)}\kgs = 0$, it is immediate that $H_{\text{open},0}\kgs = E_{0}^{(0)}\kgs$.

Doing the same for $H_{\text{open},1}$ requires a little more work due to the squaring of the Cuntz oscillators in the first term of \eqref{eq:d2cuntzopen}.
Using \eqref{eq:transformback}, one has
\beq
\label{eq:cuntzsquare}
(a_{i+1} - a_{i})^2 = A_{i+1}^{(2)} - 2A_{i}^{(1)}A_{i+1}^{(1)} + A_{i}^{(2)} + A_{i+1}^{(1)}(2\Delta z - z_{i+1}) - A_{i}^{(1)}(2\Delta z + z_i) + (\Delta z)^2
\eeq
The second term, on the hand, combines nicely due to the linear nature of the $z_i^0$:
\beq
(a_{i+1}^\dag - 2a_{i}^\dag + a_{i-1}^\dag)P_{0i}(a_{i+1} - 2a_{i} + a_{i-1}) = %
(A_{i+1}^{(1)\dag} - 2A_{i}^{(1)\dag} + A_{i-1}^{(1)\dag})P_{0i}(A_{i+1}^{(1)} - 2A_{i}^{(1)} + A_{i-1}^{(1)})
\eeq
Thus \eqref{eq:d2cuntzopen} in the site operator basis becomes
\begin{align}
& \hspace{-0.2cm} H_{\text{open},1} \nonumber \\
&= -\frac{1}{8}\sum_{i=0}^k %
(A_{i+1}^{(2)\dag} - 2A_{i}^{(1)}A_{i+1}^{(1)\dag} + A_{i}^{(2)\dag} + A_{i+1}^{(1)\dag}(2\Delta \zbb - \zbb_{i+1}) - A_{i}^{(1)\dag}(2\Delta \zbb + \zbb_i) + (\Delta \zbb)^2) \nonumber \\
&\quad \hphantom{-\frac{1}{8}\sum_{i=0}^k} \times (A_{i+1}^{(2)} - 2A_{i}^{(1)}A_{i+1}^{(1)} + A_{i}^{(2)} + A_{i+1}^{(1)}(2\Delta z - z_{i+1}) - A_{i}^{(1)}(2\Delta z + z_i) + (\Delta z)^2) \nonumber \\
\label{eq:cuntz1site}
&\quad -\frac{1}{8}\sum_{i=1}^k (A_{i+1}^{(1)\dag} - 2A_{i}^{(1)\dag} + A_{i-1}^{(1)\dag})P_{0i}(A_{i+1}^{(1)} - 2A_{i}^{(1)} + A_{i-1}^{(1)})
\end{align}

We provide some examples to further show the utility of the site operators.
First we consider the system with one site as solved in \cite{Berenstein:2006qk} for the case $\xi,~\xitt\in\mathbb{R}$ with their $\alpha = -z_1^0 = -(\xi + \xitt) / 2$.
The states for complex giant graviton coordinates are given by
\beq
\label{eq:onesiteexcited}
\ket{\Psi(p)} = \sum_{n=1}^\infty \sin(pn)e^{in\Arg(z_1^0)}A_{1}^{(n)\dag}\ket{0}_1, %
\quad E_1(p) = (1 + |z_1^0|^2 - 2|z_1^0|\cos(p))
\eeq
where $p \in [0, \pi]$ is a quasi-momentum.
The one site states take the form of a Bethe ansatz for a semi-infinite lattice and so we conclude some form of integrability.

Also given in \cite{Berenstein:2006qk} was the solution to the two site chain when $\xi = \xitt = 0$ which also happens to take the form of a Bethe ansatz.
Another special case to consider is $\xi = \xitt$.
This implies that the string collective coordinates are equal as well and there is only one free parameter $z \equiv \xi = z_1^0 = z_2^0 = \xitt$.
The site operators will produce factors of $z$ when acting on states
\begin{align}
A_i^{(1)}(A_i^{(n)\dag}\ket{0}_k) &= (A_i^{(n-1)\dag} - z A_i^{(n)\dag})\ket{0}_k \\
A_i^{(1)\dag}(A_i^{(n)\dag}\ket{0}_k) &= (A_i^{(n+1)\dag} - z A_i^{(n)\dag})\ket{0}_k
\end{align}
Even though the particle number becomes indefinite under these operations, the same factor $z$ is produced at all sites.
Thus for some states, it should be easier to relate the amplitudes of different particle excitations.
As a concrete example we consider the eigenstate
\beq
\label{eq:twositedeformed}
\ket{1, \pi/3}_z = \sum_{n=1}^\infty (-z)^{n-1}(A_1^{(n)\dag} - A_2^{(n)\dag})\ket{0}_2, \quad E_{2,z}(1, \pi/3) = \frac{3}{2}(1 + |z|^2)
\eeq
This state is the deformation of the two site excited state with one excitation and quasi-momentum $\pi/3$, which is obtained as we take $z\to 0$
\footnote{Take $n=1$ and $k=\pi/3$ in (4.11) of \cite{Berenstein:2006qk}.}.
Note that this state does not contain terms with particles excited at site one and site two.
The cancellation happens that makes this possible is
\begin{align}
&\hspace{-0.5cm} A_2^{(1)\dag}(A_2^{(1)} - \tfrac{1}{2}A_1^{(1)})\sum_{n=1}^\infty (-z)^{n-1}A_1^{(n)\dag}\ket{0}_2 \\
&= A_2^{(1)\dag}((-z) - \tfrac{1}{2}A_1^{(1)})\sum_{n=1}^\infty (-z)^{n-1}(A_1^{(n)\dag}\ket{0}_2) \\
&= A_2^{(1)\dag}\lp \sum_{n=1}^\infty (-z)^{n}A_1^{(n)\dag} - \frac{1}{2}\sum_{n=1}^\infty (-z)^{n-1}A_1^{(n-1)\dag} - \frac{1}{2}\sum_{n=1}^\infty (-z)^{n}A_1^{(n)\dag}\rp\ket{0}_2 \\
&= -\frac{1}{2}A_2^{(1)\dag}
\end{align}
There is another term with $1\leftrightarrow 2$ that allows \eqref{eq:twositedeformed} to be an exact eigenstate.
Similar cancellations occur where $\sum_{n_1=1}^\infty (-z)^{n-1} A_1^{(n_1)\dag}A_{2}^{(n_2)\dag}$ will only generate terms $A_1^{(m_1)\dag}A_{2}^{(m_2)\dag}$ with $m_1,~m_2 < n_2$.
Summing states of this type and solving the constraints for low particle number might lead to new exact eigenstates although their exact form is beyond the scope of this work.

As for integrability, we note that the space of states characterized by the site operators $A_1^{(n_1)\dag}A_{2}^{(n_2)\dag}$ can be parametrized by the sum and difference of $n_1$ and $n_2$.
When $\xi = \xitt = 0$, the space factorizes with $n_1+n_2$ being a conserved number and a Bethe ansatz exists for the subspace parametrized by $n_1 - n_2$, regardless of the number of sites.
When $\xi\neq 0$ and $\xitt\neq 0$, this is no longer the case.
It is not clear one could implement a Bethe ansatz which also takes into account the dimension parametrized by $n_1 + n_2$ as well.
The state $\ket{1, \pi/3}_{z}$ could be interpreted as a Bethe asantz with complex momentum $\log(-z)$, but at the same time looks like a Cuntz oscillator coherent state.
In particular we have that
\beq
(a_1 + a_2)\ket{1, \pi/3}_z = (-z)\ket{1, \pi/3}_z
\eeq

We note that the eigenstate $\ket{1,\pi/3}_z$ is antisymmetric under the interchange $1\leftrightarrow 2$.
There is another state that is symmetric under $1\leftrightarrow 2$ which is the deformation of the second eigenstate with $n=1$ at $\xi=\xitt = 0$.
We found numerically that the deformed symmetric eigenstate has particle excitations at both sites simultaneously without any kind of truncation occurring.
It is not obvious that this state is the result of a Bethe ansatz and so it may just be the case that a Bethe ansatz describes only a subset of the states.

The exact solvability of the one site Cuntz chain is intriguing as it provides a means by which to diagonalize the operator $A_i^{(1)\dag}A_{i}^{(1)}$.
We define the excitation operators
\beq
\Ocal_i(p) =  \sum_{n=1}^\infty \sin(pn)e^{-in\Arg(z_1^0)}P_{0i}A_{i}^{(n)}
\eeq
They satisfy the relation
\beq
A_i^{(1)\dag}A_{i}^{(1)}\Ocal_{j}(p)^\dag\kgs = \delta_{ij} E_1(p) \Ocal_{j}(p)^\dag\kgs
\eeq
The hopping terms in the one-loop Cuntz Hamiltonian ruin the possibility of writing the excited states terms of products of the operators $\Ocal_i(p_i)^\dag$ for fixed momenta $p_i$.
Instead we can take wave packets of products of the $\Ocal_i(p_i)^\dag$.

The most general form of a state for the two site Cuntz chain is
\beq
\label{eq:phiansatz}
\ket{\Phi} = \lp\sum_{i=1}^2\int dp_i\, \Phi_i(p_i)\Ocal_i(p_i)^\dag + \iint dp_1\,dp_2\, \Phi_{12}(p_1,p_2)\Ocal_1(p_1)^\dag\Ocal_2(p_2)^\dag\rp \kgs
\eeq
where the integrals run from $0$ to $\pi$ and $\Phi_1,~\Phi_2,~\Phi_{12}$ are wave functions.
Since the coefficients in the operators $\Ocal_i(p_i)$ are sine functions, the integrals produce the coefficients of the sine Fourier series of the functions $\Phi_1$, $\Phi_2$, and $\Phi_{12}$. 
In general we can take each of the $\Phi$ functions to be a sine Fourier series.
Given a state, the coefficients are just that of the individual $\prod_j A_{i_j}^{(n)\dag}P_{0i_j}$.
In particular, the wave functions for the state \eqref{eq:twositedeformed} are
\begin{align}
\Phi_1(p_1) &= \frac{2}{\pi} \sum_{n=1}^\infty (-z)^{n-1}\sin(p_1n)e^{-in\Arg(z)} = \frac{2}{\pi}\frac{\sin(p_1)}{1 + |z|^2 + 2|z|\cos(p_1)}e^{-i\Arg(z)} \\
\Phi_2(p_2) &= -\frac{2}{\pi}\frac{\sin(p_2)}{1 + |z|^2 + 2|z|\cos(p_2)}e^{-i\Arg(z)}, %
\quad \Phi_{12}(p_1, p_2) = 0
\end{align}
This allows us to directly see the distribution of momentum excitations in the excited state $\ket{1,\pi/3}_z$.
As we move the giant gravitons closer to the edge of the disk, $z\to 1$ and $\Phi_1(p_1)\propto \tan(p_1/2)$.
Higher momentum modes dominate when the giant gravitons are closer to the edge of the disk.
If we had the dual strings to the one site excited states, then we would be able to see how those strings make up a string with higher momentum.


\section{Correcting the Ground State}
\label{sec:ptheory}

Although the full spectrum of the open Cuntz Hamiltonian is out of our hands for now, one can still find the first order correction to the ground state $\kgs$.
We begin with the expansions of the full Hamiltonian $H_{\text{open}}$ %
\footnote{We exclude tree level contribution and drop an overall factor of the coupling $\lambda / 4\pi^2$.}, %
the full ground state $\ket{\Omega}$, and its exact energy $E_0$,
\beq
H_{\text{open}} = \sum_{i=0}^\infty \lp \frac{\lambda}{4\pi^2}\rp^i H_{\text{open},i}, \quad \ket{\Omega} = \sum_{i=0}^\infty \lp \frac{\lambda}{4\pi^2}\rp^i \ket{\Omega^{(i)}}, \quad E_0 = \sum_{i=0}^\infty \lp \frac{\lambda}{4\pi^2}\rp^i E_{0}^{(i)}
\eeq
where $H_{\text{open},0}\kgs = E_{0}^{(0)}\kgs$.
To second order in perturbation theory, the equation $H_{\text{open}}\ket{\Omega} = E_0\ket{\Omega}$ yields
\begin{align}
&E_{0}^{(1)} = \bgs H_{\text{open},1}\kgs = -\frac{1}{8}\frac{|\xi - \xitt|^4}{(k+1)^3} \\
\label{eq:fstate}
&(H_{\text{open},0} - E_{0}^{(0)}) \ket{\Omega^{(1)}} = -(H_{\text{open},1} - E_{0}^{(1)})\kgs \\
\label{eq:senergy}
&E_{0}^{(2)} = \bgs H_{\text{open},2}\kgs - \bra{\Omega^{(1)}} (H_{\text{open},0} - E_{0}^{(0)})\ket{\Omega^{(1)}}
\end{align}

First we solve \eqref{eq:fstate} to get the first order correction to the ground state $\ket{\Omega^{(1)}}$.
We expand the right hand side in the basis of $n$-particle states by writing $H_{\text{open},1}$ in terms of the site operators, which is given by \eqref{eq:cuntz1site}.
A couple of simplifications arise.
The terms given by \eqref{eq:cuntzsquare} act on the ground state as $(a_{i+1}^\dag - a_{i}^\dag)^2\kgs = (\Delta z)^2\kgs$.
The other is that second term of \eqref{eq:cuntz1site} vanishes on the ground state since $A_{i}^{(n)}\kgs = 0$ for all $i$ and $n>0$.
The right hand side of \eqref{eq:fstate} is then 
\begin{align}
-(H_{\text{open},1} - E_{0}^{(1)})\kgs &= \frac{1}{8}(\Delta z)^2\sum_{i=0}^{k} (A_{i+1}^{(2)\dag} - 2A_{i}^{(1)\dag}A_{i+1}^{(1)\dag} + A_{i}^{(2)\dag} + A_{i+1}^{(1)\dag}(2\Delta \zbb - \zbb_{i+1}) \nonumber \\
&\phantom{= \frac{1}{8}(\Delta z)^2\sum_{i=0}^{k}}\qquad - A_{i}^{(1)\dag}(2\Delta \zbb + \zbb_i) + (\Delta \zbb)^2)\kgs \\
&= \frac{1}{4}(\Delta z)^2\lp \sum_{i=1}^k A_{i}^{(2)\dag} - \sum_{i=1}^{k-1}A_{i}^{(1)\dag}A_{i+1}^{(1)\dag} - \sum_{i=1}^k \zbb_i A_{i}^{(1)\dag} \rp\kgs
\end{align}
We expect that $H_{\text{open},1}\kgs$ contains two particle states because the nonvanishing terms contain two Cuntz oscillators.
The appearance of one particle states is caused by the site operators containing an indefinite number of Cuntz oscillators.
Admittedly, we do not have a way to brute force solve equation \eqref{eq:fstate}.
The solution was obtained using \texttt{Mathematica} by truncating the Hilbert space, solving the linear system numerically, and realizing a pattern.
The result is
\beq
\label{eq:focttgs}
\ket{\Omega^{(1)}} = \frac{1}{4}(\Delta z)^2\sum_{i=1}^k \sum_{n=2}^\infty z_i^{n-2}A_{i}^{(n)\dag}\kgs
\eeq
Regardless, the analytic proof that \eqref{eq:focttgs} solves \eqref{eq:fstate} exists and is given in Appendix \ref{sec:proofcor}.
The state $\ket{\Omega^{(1)}}$ is normalizable, implying that the ground state remains normalizable to this order in perturbation theory,
\beq
\braket{\Omega^{(1)}}{\Omega^{(1)}} = \frac{k}{16}|\Delta z|^2
\eeq
Note that when the giant gravitons are on top of each other, $\Delta z = 0$ and the ground state receives no corrections.

To finish the computation of $E_0^{(2)}$ we need the expectation values $\bgs H_{\text{open},2}\kgs$ and $\bra{\Omega^{(1)}} (H_{\text{open},0} - E_{0}^{(0)})\ket{\Omega^{(1)}}$.
The expectation value of the first order finite difference term in \eqref{eq:d2cuntzopen} is simple to compute and yields 
\beq
\frac{1}{16}(k+1)|\Delta z|^6
\eeq
The expectation value of the third order finite difference term in \eqref{eq:d2cuntzopen} vanishes due to the nature of the string collective coordinates,
$(a_{i+2} - 3a_{i+1} + 3a_{i} - a_{i-1})\kgs = (\Delta z - 2\Delta z + \Delta z)\kgs = 0$.
The expectation value of $\sum_{i} (v_a^i)^\dag M_{ab} v_{b}^i$ in \eqref{eq:d2cuntzopen} is not as simple to compute because it has no immediate interpretation in terms of finite differences.
To simplify the computation we make use of the relations $z_{i+1} = z_i + \Delta z$ and $z_{i-1} = z_{i} - \Delta z$ and write
\beq
\begin{pmatrix}
a_{i+1}a_{i+1} \\ a_{i+1}a_{i} \\ a_{i+1}a_{i-1} \\ a_{i}a_{i} \\ a_{i}a_{i-1} \\ a_{i-1}a_{i-1}
\end{pmatrix} %
\kgs = %
\begin{pmatrix}
z_{i+1}z_{i+1} \\ z_{i+1}z_{i} \\ z_{i+1}z_{i-1} \\ z_{i}z_{i} \\ z_{i}z_{i-1} \\ z_{i-1}z_{i-1}
\end{pmatrix} %
\kgs = %
\begin{pmatrix}
1 &  2 &  1 \\
1 &  1 &  0 \\
1 &  0 & -1 \\
1 &  0 &  0 \\
1 & -1 &  0 \\
1 & -2 &  1
\end{pmatrix} %
\begin{pmatrix}
(z_i)^2 \\ z_i \Delta z \\ (\Delta z)^2 
\end{pmatrix} %
\kgs
\eeq
The expectation value is now just a product of matrices localized at a single site and we have
\beq
\label{eq:middleev}
\frac{1}{16}\sum_{i=1}^k \bgs (v_{a}^{i})^\dag M_{ab} P_{0i} v_{b}^{i} \kgs = \frac{1}{16} |\Delta z|^4\sum_{i=1}^k (1 - |z_i^0|^2)
\eeq
where the factors of $(1 - |z_i^0|^2)$ come from the projectors.
Using the orthogonality of the $n$-particle state, we have for the second expectation value
\begin{align}
&\hspace{-1cm} \bra{\Omega^{(1)}} (H_{\text{open},0} - E_{0}^{(0)})\ket{\Omega^{(1)}} \nonumber \\
&= -\bra{\Omega^{(1)}} (H_{\text{open},1} - E_{0}^{(1)})\kgs \\
&= \frac{1}{4}(\Delta z)^2\bra{\Omega^{(1)}}\lp \sum_{i=1}^k A_{i}^{(2)\dag} - \sum_{i=1}^{k-1}A_{i}^{(1)\dag}A_{i+1}^{(1)\dag} - \sum_{i=1}^k \zbb_i A_{i}^{(1)\dag} \rp\kgs \\
&= \frac{1}{16}|\Delta z|^4\sum_{i=1}^k (1 - |z_i^0|^2)
\end{align}
This cancels exactly with \eqref{eq:middleev} and we have
\beq
\label{eq:soctte}
E_{0}^{(2)} = \frac{1}{16}(k+1)|\Delta z|^6 = \frac{1}{16} \frac{|\xi - \xitt|^6}{(k+1)^5}
\eeq

Including the tree level contribution to the energy of the open string, we have for the energy of the ground state to three-loops
\beq
E_0 = (k + 1) + \frac{1}{2}\lp\frac{\lambda}{4\pi^2}\rp \frac{|\xi - \xitt|^2}{(k+1)} - \frac{1}{8}\lp\frac{\lambda}{4\pi^2}\rp^2 \frac{|\xi - \xitt|^4}{(k+1)^3}
+ \frac{1}{16}\lp\frac{\lambda}{4\pi^2}\rp^3\frac{|\xi - \xitt|^6}{(k+1)^5} + O(\lambda^4)
\eeq
This agrees with the relativistic dispersion relation $\eqref{eq:e0gauge}$ to order $\lambda^3$ and provides additional evidence that we have found the correct open string ground state in the gauge theory.


\section{Conclusions}
\label{sec:conclusions}

We have continued the analysis of open strings in constructed in $\Ncal=4$ SYM by going to three-loop order in the $\sutwo$ sector.
Shifted Cuntz oscillators were developed to handle excitations on the open string ground state which we interpret as $n$-particle states.
The site operators also provide a uniform way by which to collect Cuntz oscillator excitations while separating out the ground state from the problem of diagonalizing the open Cuntz Hamiltonian at one-loop.
The $n$-particle states can be used to compute the first order correction to the ground state and the second order correction to the energy.
The second order correction to the energy is consistent with the relativistic dispersion relation for the dual object in the string theory.

This result is non-trivial.
At this order in perturbation theory, one is implicitly summing over the entire spectrum of the one-loop Hamiltonian.
Even though we do not have the full spectrum, and do not expect to be able to find it in full generality due to a lack of integrable structures, the calculation can be done analytically.
The spectrum must thus possess some interesting property that produce the proper cancellations to yield \eqref{eq:focttgs} and \eqref{eq:soctte}.

Another cancellation occurs at this order in perturbation theory between contributions from the three-loop and the lower loop order dilatation generators.
This is very reminiscent of integrability, where the infinite number of conserved charges are related at different orders in perturbation theory \cite{Beisert:2003tq}.
Integrability is not expected to hold in our case where the giant gravitons are non-maximal, but one can not help to wonder if additional symmetry is responsible for the cancellation, or if it is just a remnant of the residual supersymmetry preserved in this sector.

Many future directions exist as a result of this analysis.
The first is to go to the four-loop order in the gauge theory as the dilatation generator is known to this order.
The dilatation generator in the Cuntz oscillators would be interesting in its own right.
One expects the appearance of finite difference terms as well as additional terms that do not take such a form.
Their structure might give insight on how to interpret the middle term of \eqref{eq:d3cuntz}.

Given the four-loop dilatation generator in the Cuntz basis, the ground state analysis can be continued to third order in perturbation theory.
It would be interesting to see how the cancellation occurs to yield the next order in the relativistic dispersion relation.
In particular, the transcendental terms must yield no contribution.

In this work we wrote down the three-loop open Cuntz Hamiltonian by assuming what was shown to be true at one and two-loops; that the giant gravitons are introduced as ordinary c-numbers.
In order to show this at three-loops we would need the full non-planar three-loop dilatation generator in the $\sutwo$ sector.
Extending the results of \cite{Sieg:2010tz} to non-planar graphs, this should be possible.



\acknowledgments
E. D. would like to thank David Berenstein for guidance and discussion in writing this note.
Work supported by the Department of Energy Office of Science Graduate Fellowship Program (DOE SCGF), made possible in part by the American Recovery and Reinvestment Act of 2009, administered by ORISE-ORAU under contract no. DE-AC05-06OR23100.


\appendix
\section{Translation of the Dilatation Generator}
\label{sec:trans}

Here we explain how to get the Cuntz Hamiltonians \eqref{eq:d1cuntz}, \eqref{eq:d2cuntz}, and specifically \eqref{eq:d3cuntz} from the planar dilatation generator in the basis of chiral functions \eqref{eq:d1chiral}, \eqref{eq:d2chiral}, \eqref{eq:d3chiral}.
First we show how to translate the chiral functions into the Cuntz oscillator language, and then we show how to properly sum them up to get the desired results.

Single trace operators in the $\mathfrak{su}(2)$ sector can be interpreted as cyclically identified words spelled from the letters $Y$ and $Z$.
All local interactions in this sector can be written as sums of products of the local permutation operator $P_{i,i+1}$ which swaps the $i$\textsuperscript{th} and the $(i+1)$\textsuperscript{th} letter.
On a word of length $L$, the chiral functions are defined by
\beq
\chi(a_1,a_2,\dots,a_n) = \sum_{r=1}^L \prod_{i=1}^n (P - I)_{a_i+r,a_i+r+1}
\eeq
with $I_{i, i+1}$ the identity operator on letters $i$ and $(i+1)$.

We can understand the chiral functions locally by considering their action on words not cyclically identified whose length is equal to the range of the interaction.
For example, the range of $\chi(1)$ is two.
It acts non-trivially only on the letter combinations $YZ$ and $ZY$,
\begin{align}
\chi(1) (ZY) &= YZ - ZY \\
\chi(1) (YZ) &= ZY - YZ
\end{align}
To convert this to the Cuntz language, we note that the $Y$'s lie on the boundary of sites and so we can label the sites preceding it or after it by $i$, $i+1$, etc.
We can then use the lowering Cuntz oscillators $a_{i}$ and zero projectors $P_{0i}$ to identify the number of $Z$'s in the input word.
To get the output word we use the raising Cuntz oscillators $a_{i}^\dag$ to put $Z$'s in the appropriate sites.
For the example of $\chi(1)$, let the $Z$ in $ZY$ be at site $i$.
Then the action of $\chi(1)$ is given by the Cuntz operators $(a_{i+1}^\dag - a_{i}^\dag)a_{i}$.
For $YZ$, we choose the $Z$ to sit at site $i$ and the action of $\chi(1)$ is $(a_{i}^\dag - a_{i+1}^\dag)a_{i+1}$.
Adding these contributions and summing over all the sites we have
\beq
\label{eq:chi1cuntz}
\chi(1) \rightarrow -\sum_{i=1}^k (a_{i+1}^\dag - a_{i}^\dag)(a_{i+1} - a_{i})
\eeq
Locally, the chiral functions will vanish on words which are all the same letter.
We do not have to consider cases with consecutive $Z$'s.
Chains of consecutive $Y$'s are sites with zero occupation which vanish under action of the lowering Cuntz oscillators.
Thus, the sum of the length of a word is equivalent to summing over the sites of the corresponding Cuntz chain.

For an example with projectors, we consider the action of the Hermitian combination $\chi(1,2) + \chi(2,1)$ on $YZY$.
One has
\beq
\label{eq:chi12cuntz}
(\chi(1,2) + \chi(2,1))(YZY) = -(ZYY - 2YZY + YYZ)
\eeq
In the Cuntz language this becomes
\beq
-(a_{i+1}^\dag - 2a_{i}^\dag + a_{i-1}^\dag)P_{0i}a_{i}
\eeq
The full expression for $\chi(1,2) + \chi(2,1)$ is
\begin{align}
\chi(1,2) + \chi(2,1)\rightarrow -\sum_{i=1}^k & (a_{i}^\dag - a_{i-1}^\dag)P_{0i}(a_{i+1} - a_{i}) + (a_{i+1}^\dag - a_{i}^\dag)P_{0i}(a_{i} - a_{i-1}) \nonumber \\
& + (a_{i+1}^\dag - a_{i}^\dag)(a_{i+1}^\dag a_{i} + a_{i}^\dag a_{i+1})(a_{i+1} - a_{i})
\end{align}

The remaining functions to be translated, $\chi(1,2,3) + \chi(3,2,1)$ and $\chi(1,3)$, act on four letter words.
Their translation is given by
\begin{align}
& \hspace{-2cm} \chi(1,2,3) + \chi(3,2,1) \rightarrow \nonumber \\
-\sum_{i=1}^k & (a_{i+2}^\dag - a_{i+1}^\dag)P_{0i+1}P_{0i}(a_{i} - a_{i-1})
+ (a_{i}^\dag - a_{i-1}^\dag)P_{0i+1}P_{0i}(a_{i+2} - a_{i+1}) \nonumber \\
& + (a_{i+1}^\dag - a_{i}^\dag)(a_{i}^\dag P_{0i} a_{i+1} + a_{i-1}^\dag P_{0i} a_{i})(a_{i} - a_{i-1}) \nonumber \\
& + (a_{i}^\dag - a_{i-1}^\dag)(a_{i+1}^\dag P_{0i} a_{i} + a_{i}^\dag P_{0i} a_{i-1})(a_{i+1} - a_{i}) \nonumber \\
& + (a_{i+1}^\dag - a_{i}^\dag)(a_{i+1}^\dag a_{i+1}^\dag a_{i}a_{i} + a_{i}^\dag a_{i}^\dag a_{i+1}a_{i+1})(a_{i+1} - a_{i}) \\
& \hspace{-2cm} \chi(1,3) \rightarrow \sum_{i=1}^k (a_{i+1}^\dag - a_{i}^\dag)(a_{i}^\dag - a_{i-1}^\dag)P_{0i}(a_{i+1} - a_{i})(a_{i} - a_{i-1})
\end{align}

Since the projectors are quadratic in the Cuntz oscillators ($P_0 = [a, a^\dag]$), we see that all terms in \eqref{eq:d3cuntz} have six Cuntz oscillators.
The chiral functions $\chi(1)$ and $\chi(1,2) + \chi(2,1)$ are only quadratic and quartic in the Cuntz oscillators, respectively.
To write \eqref{eq:d3chiral} as \eqref{eq:d3cuntz}, we make use of the identity $I = a^\dag_i a_i + P_{0i}$, nesting it when necessary.
We take advantage of the known Cuntz forms of $D_1$ and $D_2$ and replacing $\chi(1)$ and $\chi(1,2) + \chi(2,1)$ with them
\beq
D_3 = -\frac{1}{16}[(\chi(1,2,3) + \chi(3,2,1)) + \chi(1,3) + 32D_2 + 4D_1]
\eeq
Using this particular combination of chiral functions, we insert the identity into $D_1$ and $D_2$ as follows.
\begin{align}
&\hspace{-1cm} 16D_3 \rightarrow \nonumber \\
\sum_{i=1}^k & \lb (a_{i+2}^\dag - a_{i+1}^\dag)P_{0i+1}P_{0i}(a_{i} - a_{i-1})
+ (a_{i}^\dag - a_{i-1}^\dag)P_{0i+1}P_{0i}(a_{i+2} - a_{i+1}) \rr \nonumber \\
& + (a_{i+1}^\dag - a_{i}^\dag)(a_{i}^\dag P_{0i} a_{i+1} + a_{i-1}^\dag P_{0i} a_{i})(a_{i} - a_{i-1}) \nonumber \\
& + (a_{i}^\dag - a_{i-1}^\dag)(a_{i+1}^\dag P_{0i} a_{i} + a_{i}^\dag P_{0i} a_{i-1})(a_{i+1} - a_{i}) \nonumber \\
& \lr + (a_{i+1}^\dag - a_{i}^\dag)(a_{i+1}^\dag a_{i+1}^\dag a_{i}a_{i} + a_{i}^\dag a_{i}^\dag a_{i+1}a_{i+1})(a_{i+1} - a_{i})\rb \nonumber \\
& -(a_{i+1}^\dag - a_{i}^\dag)(a_{i}^\dag - a_{i-1}^\dag)P_{0i}(a_{i+1} - a_{i})(a_{i} - a_{i-1}) \nonumber \\
& + \lb 2(a_{i+1}^\dag - a_i^\dag)^2(a_{i+1}^\dag a_{i+1} + P_{0i+1} + a_{i}^\dag a_{i} + P_{0i})(a_{i+1} - a_i)^2 \rr \nonumber \\
& + \lr 2(a_{i+1}^\dag - 2a_{i}^\dag + a_{i-1}^\dag)(a_{i+1}^\dag a_{i+1} + P_{0i+1} + a_{i-1}^\dag a_{i-1} + P_{0i-1})P_{0i}(a_{i+1} - 2a_{i} + a_{i-1}) \rb \nonumber \\
& - \lb (a_{i+1}^\dag - a_i^\dag)( a_i^\dag (a_i^\dag a_i + P_{0i})a_i + P_{0i}(a_{i-1}^\dag a_{i-1} + P_{0i-1}))(a_{i+1} - a_i) \rr \nonumber \\
& + \lr (a_{i+1}^\dag - a_i^\dag)(a_{i+1}^\dag(a_{i+1}^\dag a_{i+1} + P_{0i+1})a_{i+1} + P_{0i+1}(a_{i+2}^\dag a_{i+2} + P_{0i+2}))(a_{i+1} - a_i) \rb 
\end{align}
This becomes \eqref{eq:d3cuntz} by collecting all terms with the same number of projectors and shifting indices when needed.
\begin{align}
D_3 \rightarrow H_{\text{closed},2} &= \frac{1}{16}\sum_{i=1}^k (a_{i+1}^\dag - a_{i}^\dag)^3(a_{i+1} - a_{i})^3 + v_a^{i\dag} M_{ab}P_{0i}v_b^i \nonumber \\
&\qquad + (a_{i+2}^\dag - 3a_{i+1}^\dag + 3a_{i}^\dag - a_{i-1}^\dag)P_{0i+1}P_{0i}(a_{i+2} - 3a_{i+1} + 3a_{i} - a_{i-1}) \\
v_a^i &= (a_{i+1}a_{i+1},~a_{i+1}a_{i},~a_{i+1}a_{i-1},~a_{i}a_{i},~a_{i}a_{i-1},~a_{i-1}a_{i-1}) \\
M_{ab} &= %
\begin{pmatrix}
 4 & -8 &  2 &  2 &  0 &  0 \\
-8 & 15 & -3 & -3 & -1 &  0 \\
 2 & -3 &  1 &  1 & -3 &  2 \\
 2 & -3 &  1 &  1 & -3 &  2 \\
 0 & -1 & -3 & -3 & 15 & -8 \\
 0 &  0 &  2 &  2 & -8 &  4
\end{pmatrix}
\end{align}


\section{Proof of First Order Ground State Correction}
\label{sec:proofcor}

Here we show that \eqref{eq:focttgs} solves \eqref{eq:fstate}.
One has
\begin{align}
&\hspace{-1cm}(H_{\text{open},0} - E_{0}^{(0)}) \ket{\Omega^{(1)}} \nonumber \\
&= \frac{1}{8}(\Delta z)^2\sum_{i=0}^k (A_{i+1}^{(1)\dag} - A_{i}^{(1)\dag}) (A_{i+1}^{(1)} - A_{i}^{(1)}) \lp\sum_{j=1}^{k} \sum_{n=2}^\infty z_j^{n-2} A_j^{(n)\dag}\kgs \rp \\
&= -\frac{1}{8}(\Delta z)^2\sum_{i=1}^k \sum_{j=1}^{k} \sum_{n=2}^\infty z_j^{n-2} (A_{i+1}^{(1)\dag} - 2A_{i}^{(1)\dag} + A_{i-1}^{(1)\dag})A_{i} A_j^{(n)\dag}\kgs \\
&= -\frac{1}{8}(\Delta z)^2\sum_{i=1}^k \sum_{j=1}^{k} \sum_{n=2}^\infty z_j^{n-2} (A_{i+1}^{(1)\dag} - 2A_{i}^{(1)\dag} + A_{i-1}^{(1)\dag}) \nonumber \\
&\phantom{= -\frac{1}{8}(\Delta z)^2\sum_{i=1}^k \sum_{j=1}^{k} \sum_{n=2}^\infty} \times (A_j^{(n)\dag} A_{i} + \delta_{ij}A_{j}^{(n-1)\dag}P_{0j})\kgs \\
&= -\frac{1}{8}(\Delta z)^2 \sum_{j=1}^{k} \sum_{n=2}^\infty z_j^{n-2} (A_{j+1}^{(1)\dag} - 2A_{j}^{(1)\dag} + A_{j-1}^{(1)\dag})A_{j}^{(n-1)\dag}(1 - z_j a^{\dag}_j)\kgs \\
&= -\frac{1}{8}(\Delta z)^2 \sum_{j=1}^{k} \lp \sum_{n=1}^\infty z_j^{n-1} (A_{j+1}^{(1)\dag} - 2A_{j}^{(1)\dag} + A_{j-1}^{(1)\dag})A_{j}^{(n)\dag}\ket{\Omega^{(0)}} \rr \nonumber \\
&\phantom{= -\frac{1}{8}(\Delta z)^2 \sum_{j=1}^{k} \lp \vphantom{\sum_{n=1}^\infty}\rr} %
\lr - \sum_{n=2}^\infty z_j^{n-1} (A_{j+1}^{(1)\dag} - 2A_{j}^{(1)\dag} + A_{j-1}^{(1)\dag})A_{j}^{(n)\dag}\ket{\Omega^{(0)}}\rp \\
\label{eq:midproofb6}
&= -\frac{1}{8}(\Delta z)^2 \sum_{j=1}^{k} (A_{j+1}^{(1)\dag} - 2A_{j}^{(1)\dag} + A_{j-1}^{(1)\dag})A_{j}^{(1)\dag}\kgs \\
\label{eq:midproofb7}
&= \frac{1}{4}(\Delta z)^2 \lp \sum_{j=1}^{k} A_j^{(2)\dag} - \sum_{j=1}^{k-1} A_{j}^{(1)\dag}A_{j+1}^{(1)\dag} - \sum_{j=1}^k \zbb_j A_{j}^{(1)\dag} \rp\kgs \\
&= -(H_{\text{open},1} - E_{1}^{(1)}) \kgs
\end{align}
To go from \eqref{eq:midproofb6} to \eqref{eq:midproofb7} we used the relation $A_{j}^{(1)\dag}A_{j}^{(1)\dag} = A_{j}^{(2)\dag} - \zbb A_{j}^{(1)\dag}$.


\bibliographystyle{jhep}
\bibliography{ref}

\end{document}